\documentclass[english,aps,pra,twocolumn,showpacs,superscriptaddress,floatfix]{revtex4-1}
\usepackage[T1]{fontenc}
\usepackage[latin1]{inputenc}
\usepackage{amsmath}
\usepackage{graphicx}
\usepackage{amssymb}
\usepackage{dcolumn} 
\usepackage{bm} 

\makeatletter

\usepackage{amsfonts}
\usepackage{epsfig}
\usepackage{dsfont}

\newcommand{\Eq}[1]{Eq.~(\ref{#1})}

\newcommand{\be}{\begin{equation}}
\newcommand{\bea}{\begin{eqnarray}}
\newcommand{\eea}{\end{eqnarray}}
\newcommand{\ee}{\end{equation}}

\newcommand{\quot}[1]{``#1''}

\newcommand{\A}{{\cal A}}

\newcommand{\Q}{{\cal Q}}
\newcommand{\C}{{\cal C}}
\newcommand{\F}{{\cal F}}
\newcommand{\Pb}{{\cal P}}
\newcommand{\U}{{\cal U}}
\newcommand{\Sc}{{\cal S}}
\newcommand{\V}{{\cal V}}
\newcommand{\M}{{\cal M}}

\def\H{{\cal H}}

\usepackage{babel}
\makeatother

\begin{document}

\title{Sequential quantum cloning under real-life conditions}

\author{Hamed \surname {Saberi}}
\affiliation{Department of Physics, Shahid Beheshti University, G.C., Evin, Tehran 19839, Iran}
\affiliation{Institute for Theoretical Physics, University of Regensburg, 93040 Regensburg, Germany}
\affiliation{Physics Department, Arnold Sommerfeld Center for Theoretical Physics, and Center for NanoScience, Ludwig-Maximilians-Universit\"{a}t, Theresienstr.~37, 80333 Munich, Germany}

\author{Yousof \surname {Mardoukhi}}
\affiliation{Department of Physics, Shahid Beheshti University, G.C., Evin, Tehran 19839, Iran}


\date{April 11, 2012}

\begin{abstract}

  We consider a sequential implementation of the optimal quantum cloning machine of Gisin and Massar and propose optimization protocols
  for experimental realization of such a quantum cloner subject to the real-life restrictions. We demonstrate how exploiting the matrix-product state (MPS) formalism and the ensuing variational optimization techniques reveals the intriguing algebraic structure of the Gisin-Massar output of the cloning procedure and brings about significant improvements to the optimality of the \emph{sequential} cloning prescription of Delgado \emph{et al} [Phys. Rev. Lett. \textbf{98}, 150502 (2007)]. Our numerical results show that the orthodox paradigm of optimal quantum cloning can in practice be realized in a much more economical manner by utilizing a considerably lesser amount of informational and numerical resources than hitherto estimated. 
  Instead of the previously predicted linear scaling of the required ancilla dimension $D$ with the number of qubits $n$, our recipe allows a realization of such a sequential cloning setup with an experimentally manageable ancilla of dimension at most $D=3$ up to $n=15$ qubits.
  We also address satisfactorily the possibility of providing an optimal range of sequential ancilla-qubit interactions for optimal cloning of arbitrary states under realistic experimental circumstances when only a restricted class of such bipartite interactions can be engineered in practice.

\end{abstract}

\pacs{03.67.Lx; 03.67.Bg; 03.65.Ta; 02.70.-c; 71.27.+a}
 	

\maketitle

\section{Introduction}
\label{sec:intro}

The superposition principle of quantum mechanics precludes quantum information from being perfectly \quot{cloned} and transformed into classical information. Such a fundamental impossibility and chief difference between classical and quantum information is formally phrased in terms of the so-called \quot{no-cloning} theorem~\cite{Wootters1982} of quantum computation and quantum information~\cite{Nielsen2000}. Nonetheless, quantum cloning machines (QCM) allow copying of arbitrary and \emph{a priori} unknown quantum states with an imperfect albeit optimal \emph{fidelity} as a measure of the quality of copies (clones)~\cite{Buzek1996,Gisin1997,Scarani2005}. Such QCMs provide an optimal way of isometrically (unitarily) evolving an initial set of $N$ input qubits all in the same unknown state (augmented by $M-N$ blank qubits) into the final state of $M$ approximate clones, with the whole evolution described by the map $N \to M$ ($M \to M$).
However, engineering such maps through a single application of a \emph{global} isometry (unitary) operation that is capable of entangling all input qubits simultaneously is in general prohibitively difficult to realize in an experiment. As a consequence, all experimental realizations of QCMs hitherto either in optical systems~\cite{Huang2001,Lamas2002,Martini2002,Fasel2002,Zhao2005,Bartuskova2007,Nagali2009} or NMR setups~\cite{Cummins2002,Du2005,Chen2007} have suffered from the requirement for exceedingly high experimental demands that particularly poses major obstacles in scaling the devices to multiqubits. In an effort to work around such an experimental challenge, a \emph{sequential} prescription for an experimentally manageable implementation of the multiqubit entangler of the optimal QCMs has been proposed by Delgado \emph{et al}~\cite{Delgado2007}.
\begin{figure}[b]
\centering
\includegraphics[width=1\linewidth]{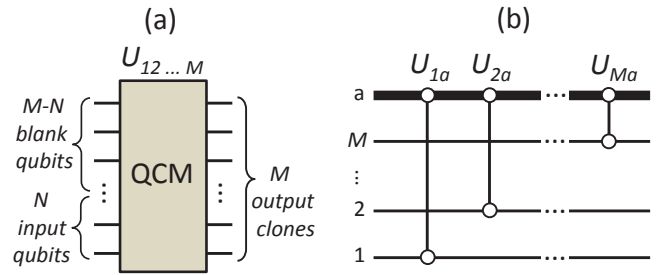}
\caption[Sequential quantum cloning machine]{$M \to M$ quantum cloning machine implemented (a) globally and (b) sequentially with the aid of an ancillary system. Each bipartite unitary $U_{ka}$ in (b) acts \emph{only} on the Hilbert space of qubit $k$ and ancilla $a$ and leaves other qubits intact. The fact has been illustrated by using vertical lines with open circles at each qubit they act upon. Ancilla degrees of freedom are shown by solid bold lines throughout.}
\label{seq_cloning}
\end{figure}
In sequential quantum cloning paradigm of Delagdo \emph{et al} an ancillary system is introduced to interact sequentially and only once with each qubit in a row and is set to decouple from the qubit chain in the last step~\cite{Delgado2007}. This is essentially equivalent to a nontrivial decomposition of the multiqubit entangling operation of the QCM into a one-way sequence of consecutive two-body ancilla-qubit operations (see Fig.~\ref{seq_cloning}). Although a general \emph{genuinely} entangling multiqubit operation cannot be implemented via such a sequential prescription~\cite{Lamata2008,Saberi2011}, the case of sequential implementation of an $N \to M$ isometry required for QCM tasks was proved to be always possible~\cite{Lamata2008}.

From experimental point of view, various physical settings employ sequential operations for generation of entangled multiqubit states. Paradigmatic setups include the photonic streams of cavity or circuit QED with a cavity mode as the ancillary system~\cite{Kuhn2002,McKeever2004,Schoen2005,Schoen2005_2,Schoen2007,Wilk2007,Weber2009,Saberi2009} and the sequential generation of electron-spin entangled states in quantum dot setups utilizing a completely mixed nuclear spin bath as the ancilla~\cite{Christ2008}. In terms of the practical benefits, sequentiality and the absence of measurements guaranteed by a unitarily decoupling ancilla may tremendously facilitate the physical implementation of a generically complex global cloning operation.

Furthermore, from theoretical perspective, it has been demonstrated~\cite{Delgado2007} that the multipartite entangled state of the output clones produced by a sequentially implemented quantum cloner can be characterized in terms of the hierarchy of the so-called matrix-product states (MPS)~\cite{Perez2007} that had previously arisen in the context of spin chain models~\cite{Fannes1992} and are by now recognized as a unifying framework for rephrasing and optimizing various numerical renormalization group techniques of strongly correlated systems~\cite{Saberi2008,Weichselbaum2009,Pizorn2012}. An MPS representation of an $n$-qubit output of a QCM is then given by
\begin{eqnarray}
\label{eq:MPS_nqubit}
|\psi_{\mathrm{out}}\rangle=\sum_{i_{n} \dots i_1=0}^1 \langle\varphi_F| V_{[n]}^{i_{n}} \dots V_{[1]}^{i_1} |\varphi_I\rangle
|i_{n},\dots,i_1\rangle   \; ,
\end{eqnarray}
where $D_k\times D_{k+1}$-dimensional matrix $V_{[k]}^{i_k}$ represents the physical interaction between ancilla and $k$th qubit with local space $|i_k\rangle$, and with $|\varphi_I\rangle$ and $|\varphi_F\rangle$ denoting the initial and final ancilla state, respectively. The \emph{bond dimension} of an MPS is defined as $D \equiv \mathrm{max}_k D_k$.

As pointed out by Delgado \emph{et al}, it turns out that the minimal bond dimension of the MPS representation of (\ref{eq:MPS_nqubit}) coincides the required ancilla dimension $D$ (e.g., the number of atomic levels) for sequential realization of the desired quantum cloner~\cite{Delgado2007}. Thus, apart from the algorithmic facilitation of the sequential prescription in terms of the decomposition of a global cloner into consecutive two-body interactions, its practical feasibility and scalability yet hinges crucially upon ancilla demands and the way it scales with the number of clone qubits. A detailed analysis of such an issue revealed that the required minimal ancilla dimension $D$ scales only \emph{linearly} with the number of clones $M$ (more precisely $D=2M$ for the case of universal symmetric cloning of Gisin-Massar as described in the subsequent section). This result promises an experimentally affordable ancilla-assisted scenario for quantum cloning of a \emph{small} number of qubits, but yet infers the one that demands more and more experimental resources upon increasing the number of qubits. On the other hand, cloning of arbitrary unknown states entails access to a vast variety of ancilla-qubit interactions ($V_{[k]}^{i_k}$ in \Eq{eq:MPS_nqubit}), whereas some of the required ones may be unattainable within a given physical setup. In this sense, an important experimental question will be if the existing physical setups could still fulfill the resource requirements of the proposed sequential QCM (SQCM). And if the answer is no, to what extent the numerical resources of the SQCM algorithm can be further optimized toward less costly experimental demands and, in turn, allowing realization of SQCM under such restricted experimental conditions. It is the purpose of the present paper to address these issues.

This paper is organized as follows: Sec.~\ref{sec:GM_MPS} sets the scene by rephrasing the optimal cloning machine of Gisin and Massar in terms of MPS and providing some technical details on simulation of such a cloner. In Sec.~\ref{sec:D_reg} the issue of restrictions on the ancilla dimension $D$ is addressed and numerical \quot{regularization} techniques for achieving the cloning task with the minimal possible resources are presented. Section~\ref{sec:reg_anc_qubit} deals with restrictions on ancilla-qubit interactions and optimization protocols for realizing the cloning task with the minimal accessible class of physical operations. Finally, Sec.~\ref{sec:conclusions} contains our conclusions and an assessment of the applicability of our proposed methods to future realization of \emph{scalable} quantum cloning.

\section{Matrix-product representation of the Gisin-Massar state}
\label{sec:GM_MPS}

We consider here the universal symmetric cloning~\cite{Buzek1996,Gisin1997} of a single qubit in an arbitrary unknown state
$|\psi\rangle= \alpha |0\rangle + \beta |1\rangle$ to $M$ clones described by the isometry map $1 \to M$
\begin{eqnarray}
\label{eq:Gisin_Massar}
\nonumber
|\psi\rangle \to |\psi_{\mathrm{out}}\rangle & = & |GM_M(\psi)\rangle\\
\nonumber
& \equiv & \sum_{j=0}^{M-1} \alpha_j |(M-j) \psi, j\psi^{\bot}\rangle_S \\
& &  \qquad \otimes  |(M-j-1) \psi^*, j \psi^{*\bot}\rangle_S   \; ,
\end{eqnarray}
with the coefficients $\alpha_j= \sqrt{{2(M-j) \over M(M+1)}}$, and $|(M-j) \phi, j\phi^{\bot}\rangle_S$ denotes the normalized completely symmetric state with $M-j$ qubits in state $|\phi\rangle$ and $j$ qubits in the orthogonal state $|\phi^{\bot}\rangle$. The resulting multiqubit state from the cloning procedure, the so-called Gisin-Massar state, describes an entangled state of $M$ clones augmented by $M-1$ anticlones that are introduced to guarantee the optimality of the cloning procedure. We shall later exploit such a \quot{parity} feature of clones and anticlones for encoding $\alpha_j$ into MPS representation of the Gisin-Massar state.

\begin{figure}[t]
\centering
\includegraphics[width=1\linewidth]{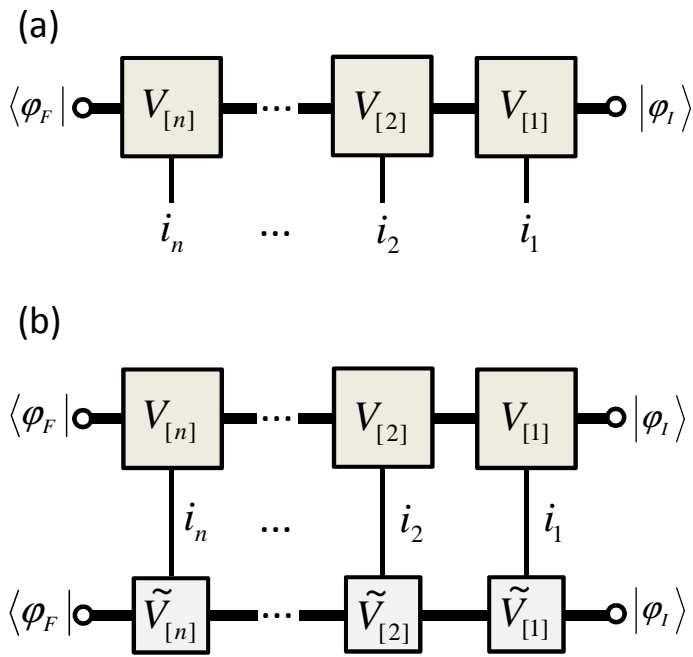}
\caption[MPS representation of Gisin-Massar]{(a) Graphical representation of the MPS description of Gisin-Massar state and (b) the overlap of states used to calculate the contractions in the cost function~\Eq{eq:cloning_var}. The boxes represent $V_{[k]}$ matrices of a sequentially prepared state of Gisin-Massar form. The links connecting the boxes represent indices that are being contracted (or summed over). The more information (e.g. larger ancilla dimension or larger number of ancilla-qubit couplings) the matrices contain, the darker and larger the associated boxes are drawn throughout.}
\label{MPS_GM}
\end{figure}

An $(2M-1)$-qubit state of Gisin-Massar written in computational basis of the form
\begin{eqnarray}
\label{eq:GM_comp_basis}
\nonumber
\hspace{-5mm}|GM_M(\psi)\rangle = \sum_{i_{2M-1}=0}^1 \sum_{i_{2M-2}=0}^1 \dots \sum_{i_1=0}^1
c^{\psi}_{i_{2M-1}, i_{2M-2}, \dots ,i_1}\\
|i_{2M-1}, i_{2M-2}, \dots ,i_1\rangle   \; ,
\end{eqnarray}
can be cast into a canonical MPS representation with minimal bond dimension $D$~\cite{Schoen2005,Schoen2007}. The symmetrization requirement of the Gisin-Massar state within either clone or anticlone subspace establishes the algebraic connection between $\alpha_j$ and $c^{\psi}$. It indeed makes most of the coefficients $c^{\psi}$ vanish or to be identical.
Distinct values of $\alpha_j$ correspond only to \emph{a priori} symmetrized products of clone kets and the anticlone ones and are equal in number to the number of clones $M$ to be produced. We shall see in the subsequent section that this feature profoundly affects the simulability of such a state in terms of the required numerical resources.

The first step in achieving an MPS representation of the Gisin-Massar state is to perform Vidal decomposition~\cite{Vidal2003} on (\ref{eq:GM_comp_basis}). Treating a bipartite decomposition of the coefficient $\C \equiv c^{\psi}_{i_{2M-1}| i_{2M-2}, \dots ,i_1}$ as a matrix with indices $i_{2M-1}$ and the combined one $\tilde{i}_{2M-1} \equiv i_{2M-2}, \dots ,i_1$, we perform singular value decomposition (SVD)~\cite{Golub1996,Horn1991} on $\C$
\begin{eqnarray}
\label{eq:SVD1_C}
\C=\U \Sc \V^{\dagger}   \; .
\end{eqnarray}
Identifying now the left unitary as $\U \equiv V_{[2M-1]}^{i_{2M-1}}$, we proceed with the singular value decomposition of
the remaining part $\M^{[2M-1]} \equiv \Sc \V^{\dagger}$ within the next bipartite decomposition $i_{2M-1}, i_{2M-2} | i_{2M-3}, \dots, i_1$ and obtain
\begin{eqnarray}
\label{eq:SVD2_C}
\M^{[2M-1]}=\U' \Sc' \V'^{\dagger}   \; .
\end{eqnarray}
Identifying again the left unitary as $\U' \equiv V_{[2M-2]}^{i_{2M-2}}$ and iterating such a procedure till the end of the chain yields an open-boundary MPS representation of the Gisin-Massar state of the form (\ref{eq:MPS_nqubit}) with $n$ replaced by $2M-1$.
A useful graphical representation of Gisin-Massar state in MPS form has been shown in Fig.~\ref{MPS_GM}(a).

Some remarks on technical details of reshaping the $c^{\psi}$ (or equivalently $\alpha_j$s) into $V$-matrices are in order. An operational ambiguity may arise from contributions degenerate in $\alpha_j$ coefficient, though, belonging to cloning of $|0\rangle$ or $|1\rangle$
\begin{subequations}
\label{eq:GM_parity}
\begin{eqnarray}
\nonumber
|GM_M(0)\rangle = \sum_{j=0}^{M-1} \alpha_j|(M-j)0,j1\rangle_{S} \\
\otimes |(M-j-1)1,j0\rangle_{S} , \\
\nonumber
|GM_M(1)\rangle = \sum_{j=0}^{M-1} \alpha_j|(M-j)1,j0\rangle_{S} \\
\otimes |(M-j-1)0,j1\rangle_{S} , \;
\end{eqnarray}
\end{subequations}
whose linear combinations constitute the expression \Eq{eq:Gisin_Massar} for cloning of an arbitrary state $|\psi\rangle$. It will be, however, essential to distinguish such degenerate contributions in each bipartite combination of the indices in order to realize an \emph{in situ} restructuring of the information. For this purpose, we suggest a parity-bit encoding scheme to get around this difficulty.
The scheme relies on the basic observation that contributions associated with cloning of $|0\rangle$ ($|1\rangle$) enjoy even (odd) parity in terms of the number of qubits in state $|1\rangle$. Hence, such a degeneracy is formally lifted upon the action of a parity-bit operator $\Pb$ with eigenvalues $\pm 1$ for a register of qubits with even (odd) parity. The technical details of the latter procedure go beyond the scope of the present paper and will be published separately.
The parity bit encoding scheme has already been employed in the context of classical and quantum cryptography~\cite{Bennett1996}, redundant array of independent disks (RAIDs) and optical quantum computing schemes~\cite{Hayes2010}.

Following such a constructive recipe, results in a Gisin-Massar state in canonical form with the maximal bond dimension $D=d^{\lfloor n/2 \rfloor}$ where $d$ denotes the local dimension ($d=2$ for qubits)~\cite{Perez2007}. Such a maximal bond dimension will be the minimal required one for the canonical description of the target state, too, provided that all the constituent $V$ matrices are full-rank, otherwise rank-deficient matrices could give rise to a bond dimension lower than the canonical one. In this sense, as we shall see in the subsequent section, rephrasing Gisin-Massar in terms of MPS allows to obtain vivid insights into the resource requirements of the Gisin-Massar cloning machine.

\section{Regularization of the ancilla dimension}
\label{sec:D_reg}

\begin{figure*}[t]
\centering
\includegraphics[width=1\linewidth]{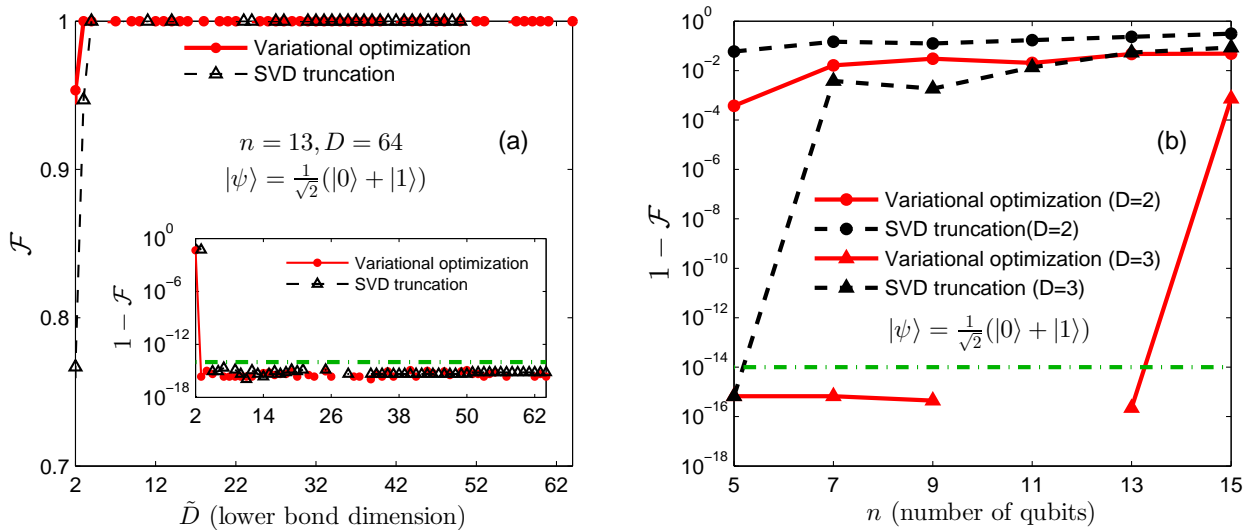}
\caption[Results for regularization of the ancilla dimension]{(Color online) Results for the regularization of the ancilla dimension $D$. The inset of (a) shows the error (1-$\F$) on a log scale which turn out to be of the order of the numerical noise (depicted by the dash-dotted green line) beyond $\tilde{D} > 2$, thereby confirming the regularizability of the ancilla dimension for a state of Gisin-Massar form. The missing data points in (a) and (b) reflect calculations beyond the lower bound of the numerical precision.}
\label{D_reg}
\end{figure*}

The analytical study of Delgado \emph{et al} already clarified that a multiqubit state of the Gisin-Massar form exhibits a rank deficient MPS representation with linear scaling of the bond dimension $M$ with the number of clones. Nevertheless, the resource costs of the result yet seems unaffordable for large number of qubits. One may then wonder if any optimization protocol can be devised to allow the accomplishment of the same task with lesser amount of resources. In other words, we pose the following question:
how well a state of Gisin-Massar form $|GM_{M}(\psi)\rangle$ with bond dimension $D$ (and equivalently requiring an ancilla of dimension $D$ for its sequential generation) can be represented if only an ancilla with a smaller number of levels, $\tilde{D}<D$, is available? More formally: given a state $|GM_{M}(\psi)\rangle$, with bond dimension $D$, what is the optimal MPS $|\widetilde{GM}_M(\psi)\rangle$ of lower bond dimension $\tilde{D}<D$ that \quot{regularizes} the original huge ancilla demands by minimizing the distance
\begin{eqnarray}
\label{eq:MPS_smaller_D}
\hspace{-4mm} \min_{\dim({|\widetilde{GM}_M (\psi)\rangle})=\tilde{D}<D  }
\parallel |GM_M(\psi)\rangle - |\widetilde{GM}_M(\psi)\rangle \parallel^{2}  \; .
\end{eqnarray}

We exploit two techniques that were developed by Saberi \emph{et al}~\cite{Saberi2009, Saberi2009_2} in the context of sequential generation of entangled states to perform the MPS approximation above, both exploiting a suitably designed local
optimization of the $V$-matrices in \Eq{eq:MPS_nqubit}. In the first approach, use is made of
a corollary of the singular value decomposition theorem from linear algebra to perform a local
optimization procedure which might be referred to as ``MPS compression'', in analogy to the image compression technique
already used in computer science and engineering~\cite{Andrews1976}. Let the SVD of matrix $A$ with $\mathrm{rank}(A)=r$ be
given by
\begin{eqnarray}
\label{eq:svd_A}
A=\sum_{i=1}^{r} \sigma_i u_i v_i^{\dagger} \; .
\end{eqnarray}
Then, the best possible lower-rank approximation to $A$ that minimizes the Frobenius-norm distance~\cite{frob}
\begin{eqnarray}
\label{eq:fro_dist_A}
\min_{\mathrm{rank}(\tilde{A})=\tilde{r}<r} {\|A-\tilde{A}\|}_{F}  \; ,
\end{eqnarray}
is given by~\cite{Golub1996,Horn1991}
\begin{eqnarray}
\label{eq:best_fro}
\tilde{A}=\sum_{i=1}^{\tilde{r}} \sigma_i u_i v_i^{\dagger}  \; .
\end{eqnarray}
The latter indicates a \emph{truncation} scheme in which one retains only the $\tilde{r}$ largest singular values of $A$ to form the desired optimal lower-rank matrix $\tilde{A}$ and discards the rest.
Applying the outlined truncation scheme to each matrix $V_{[k]}^{i_k}$, $k=1, \dots, n$, in \Eq{eq:MPS_nqubit}, yields an MPS of lower bond dimension $\tilde{D}=D-(r-\tilde{r})$ which is locally closest to the original MPS of bond dimension $D$ in the Frobenius norm sense.
It is noteworthy to mention that the precision of the method hinges to a great extent upon how well-decaying the singular values spectrum of the underlying matrices are.

\begin{table*}[!t]
\caption{Results for regularization of the ancilla-qubit interactions}
\centering
\begin{tabular}{c c c c c c}
\hline\hline
number of qubits ($n$) &3 &5 &7 \\ 
\hline \\
$1-\F$ (without auxiliary unitaries) & 0.5000  & 0.4730 &  0.5684 \\  [1.5ex]
\hline \\
$1-\F$ (with auxiliary unitaries) & 8.6938$\times 10^{-9}$ & 3.7714$\times 10^{-4}$ & 1.6600$\times 10^{-2}$\\  [1.5ex]
\hline
\hline
\end{tabular}
\label{table:AQ_reg}
\end{table*}

In the second approach~\cite{Saberi2008,Saberi2009,Saberi2009_2}, a DMRG-inspired variational optimization of $V$-matrices is performed~\cite{Verstraete2004_2,Schollwock2005, White1992,Saberi2009,Saberi2009_2,Weichselbaum2009} to obtain
the best possible approximation to $|GM_M(\psi)\rangle$ in the space of all MPS of lower bond dimension $\tilde{D}<D$ of the form
\begin{eqnarray}
\label{eq:MPS_Dtilde}
\nonumber
|\widetilde{GM}_M(\psi)\rangle = \sum_{i_{n} \dots i_1=0}^1 \langle\varphi_F| \tilde{V}_{[n]}^{i_{n}} \dots \tilde{V}_{[1]}^{i_1} |\varphi_I\rangle \\
\hspace{3cm}
|i_{n},\dots,i_1\rangle   \; ,
\end{eqnarray}
by solving the minimization problem of \Eq{eq:MPS_smaller_D}
under the constant-norm constraint $\langle \widetilde{GM}_M(\psi)|\widetilde{GM}_M(\psi)\rangle=1$, which is taken care of by introducing a Lagrange multiplier $\lambda$. Varying \Eq{eq:MPS_smaller_D}
with respect to the matrices defining $|\widetilde{GM}_M(\psi)\rangle$ leads to a set of equations,
one for each $i_k$, of the form
\begin{eqnarray}
\label{eq:cloning_var}
\nonumber
\frac{\partial}{\partial \tilde{V}^{i_k}_{[k]}} \Bigl[(1+\lambda) \:
\langle \widetilde{GM}_M(\psi)|\widetilde{GM}_M(\psi)\rangle \\
- 2 {\rm Re} \bigl\lbrace\langle \widetilde{GM}_M(\psi)|\widetilde{GM}_M(\psi)\rangle \bigr\rbrace \Bigr]=0  \; ,
\end{eqnarray}
which determines the optimal $\tilde{V}$-matrices of the desired state $|\widetilde{GM}_M(\psi)\rangle$.
These equations can be solved very efficiently using a ``sweeping procedure''~\cite{Saberi2008} in which
one fixes all but the $k$'th $\tilde{V}$-matrix and solves the
corresponding Eq.~(\ref{eq:cloning_var}) for the matrix $\tilde{V}_{[k]}^{i_k}$.
Then one moves on to the neighboring site and, in this fashion, sweeps back and forth
through the chain until the convergence is reached. The cost function contains the overlap terms between the target state $|GM_M(\psi)\rangle$ and the simulational one $|\widetilde{GM}_M(\psi)\rangle$ which can be straightforwardly
calculated in MPS representation as illustrated in Fig.~\ref{MPS_GM}(b).

We have implemented numerically the two regularization methods outlined above and the results are illustrated in Fig.~\ref{D_reg}. Focusing on the cloning of an input state of the form $|\psi\rangle= \frac{1}{\sqrt{2}} (|0\rangle + |1\rangle)$, we define the \emph{fidelity} of the procedure to be the overlap between the target Gisin-Massar state and the simulational one of the form
\begin{eqnarray}
\label{eq:fidelity}
\F= \langle GM_M(\psi)|\widetilde{GM}_M(\psi)\rangle  \; .
\end{eqnarray}
Strikingly, Fig.~\ref{D_reg}(a) suggests that, for instance, a Gisin-Massar state of $n=13$ qubits (associated with $M=7$ clones) can be simulated with almost perfect fidelity ($1-\F \approx 10^{-16}$) with a regularized dimension of $D=3$, and to a lesser degree ($1-\F \approx 10^{-2}$) with $D=2$. Note that, the analytical result of Delgado \emph{et al} requires an ancilla of dimension $D=2M=14$ for realization of such a task, whereas our numerics allows the experimental realization of SQCM with an ancilla of dimension \emph{only} $D=3$ which is considered to be amenable to the existing physical setups. The performed procedure can be regarded as a kind of \quot{tomography} of Gisin-Massar state within the subspace of the constituent dimensions. As is evident from Fig.~\ref{D_reg}(a), the dimension components $\tilde{D}$ beyond $D=3$ make negligible contributions to the representation of the state of Gisin-Massar with $n=13$.

Figure~\ref{D_reg}(b) shows the scaling of the fidelity with the number of qubits $n$ using an ancilla of dimension $D=2$ and $D=3$.
Since variational optimization approach allows for the feedback of information through several sweeps, it generally performs better than SVD truncation scheme. The SVD truncation scheme is only a local optimization of information in the sense that each matrix is being optimized independently from the rest of the matrices, whereas the variational optimization, though being itself a local optimization protocol too, makes up for such a locality by performing several sweeps and allowing the feedback of entanglement through the chain.

All in all, the simulation promises the possibility of realizing SQCM with a manageable amount of ancillary levels $D=3$ up to $n=15$ qubits. We believe the regularizability of the Gisin-Massar state mainly originates from symmetrization requirements of the optimal protocol already discussed in previous section.

\section{Regularization of the ancilla-qubit interactions}
\label{sec:reg_anc_qubit}

A state of Gisin-Massar form can be generated sequentially provided that the required ancilla dimension $D$ and ancilla-qubit unitaries are available~\cite{Schoen2005,Schoen2007,Saberi2009}. However, in general, some of the required local ancilla-qubit unitaries may not be accessible to a given physical setup. Given such a relevant constrained optimization problem~\cite{Bertsekas1996}, we wonder if an optimization protocol can be devised by which a Gisin-Massar ``target'' state can be approximately generated with maximal fidelity.

In general, the unitary time evolution of the joint system ancilla-qubit at step $k$ of the sequential generation is described by a general unitary $U_{[k]}^{\A\Q}:\H_\A\otimes\H_\Q \to \H_\A\otimes\H_\Q$
\begin{subequations}
\begin{eqnarray}
U_{[k]}^{\A\Q} & = & e^{-iH_{[k]}^{\A\Q} t/\hbar} ,\\
H_{[k]}^{\A\Q} & = & \sum_{j_{\A},j_{\Q}=0}^{3} h_{j_{\A} j_{\Q}}^{[k]} \sigma_{j_{\A}} \otimes \sigma_{j_{\Q}}   \; ,
\end{eqnarray}
\end{subequations}
where $H_{[k]}^{\A\Q}$ is a general bipartite Hamiltonian that entangles the
ancilla with the $k$'th qubit, $h_{j_{\A} j_{\Q}}^{[k]}$ are real-valued coupling constants, and $\sigma_1$, $\sigma_2$,
$\sigma_3$ are the usual Pauli sigma matrices, with $\sigma_0\equiv I$ the identity matrix. For simplicity, we have considered the case $D=2$, but similar generators of SU($N$) may readily be employed for $D>2$.


As an illustrative case, now suppose that only a restricted set of unitaries produced by the experimentally realizable entangling Hamiltonian of the form of the $XXZ$-model~\cite{Brown2003} are available
\begin{eqnarray}
\label{eq:restricted_Hamiltonian}
\tilde{H}_{[k]}^{\A\Q} = h_1^{[k]} (\sigma_1 \otimes \sigma_1+\sigma_2 \otimes \sigma_2)+ h_2^{[k]} \sigma_3 \otimes \sigma_3 \; ,
\end{eqnarray}
containing two nonzero couplings $h_1^{[k]} \equiv h_{11}^{[k]}=h_{22}^{[k]}$ and $h_2^{[k]} \equiv h_{33}^{[k]}$.
Given a state of the Gisin-Massar form and the restricted entangling Hamiltonian of \Eq{eq:restricted_Hamiltonian}, the aim is to tailor the restricted unitary operations ${\tilde{U}}_{[k]}^{\A\Q}=e^{-i {\tilde{H}_{[k]}^{\A\Q}} t/\hbar}$ in such a way that when applied sequentially to
an arbitrary initial state of the joint system $|\Phi_I\rangle=|\varphi_I\rangle \otimes |\psi\rangle$, yield a state of the form
\begin{eqnarray}
\label{eq:joint_psi}
|\tilde{\Psi}\rangle=\tilde{U}_{[n]}^{\A\Q}\dots \tilde{U}_{[2]}^{\A\Q} \tilde{U}_{[1]}^{\A\Q} |\Phi_I\rangle \; ,
\end{eqnarray}
that is ``closest'' to the target state of the form $|\varphi_F \rangle \otimes | GM_M(\psi) \rangle$, where $|\varphi_F \rangle$ is arbitrary~\cite{uni_iso}.
In the ideal case, when the fidelity reaches unity, the ancilla can be set to decouple unitarily in the last step~\cite{Schoen2007}. However, this ceases to be the case in general when the allowed ancilla-qubit unitaries are restricted. Thus, the optimization problem shall be carried out with respect to the cost function of the from
\begin{eqnarray}
\label{eq:cost_func}
f[|\tilde{\Psi}\rangle]= \parallel |\tilde{\Psi}\rangle - |\varphi_F\rangle \otimes |GM_M(\psi)\rangle  \parallel^{2}  \;,
\end{eqnarray}
involving a multivariable cost function in $|\varphi_F\rangle$ and the interaction couplings $\{h_1^{[n]},h_2^{[n]}; \dots; h_1^{[1]}, h_2^{[1]}\}$, as the \emph{variational parameters}, which can be solved in an iterative procedure developed by Saberi \emph{et al} in the context of sequential generation of entangled states~\cite{Saberi2009}: One starts by picking a particular
unitary, say $\tilde{U}_{[k]}^{\A\Q}$, and minimizes the cost function in~\Eq{eq:cost_func}, varying over $\{h_1^{[k]}, h_2^{[k]}\}$ and regarding couplings of all the other unitaries as fixed. Then one moves on to the neighboring unitary and optimize its couplings. When
all unitaries have been optimized locally, one sweeps back again and so forth until convergence.
Varying over the vector $|\varphi_F\rangle$ and using the resulting optimal one, the cost function boils down to
\begin{eqnarray}
\label{eq:cost_func2}
f[|\tilde{\Psi}\rangle] = 2(1-\| \langle \tilde{\Psi} | GM_M{(\psi)} \rangle \|)  \;,
\end{eqnarray} 
suggesting the definition of the fidelity of the procedure as
\begin{eqnarray}
\label{eq:fid_AQ}
\F \equiv \| \langle \tilde{\Psi} | GM_M(\psi)\rangle \|  \; .
\end{eqnarray}

For the restricted entangling Hamiltonian of \Eq{eq:restricted_Hamiltonian}, however, the variational space is so small that the variational optimization of the restricted unitaries \emph{per se} does not result in much overlap with the target state $|GM_M(\psi)\rangle$, as illustrated in the first row of Table~\ref{table:AQ_reg}.

\begin{figure}[t]
\centering
\includegraphics[width=1\linewidth]{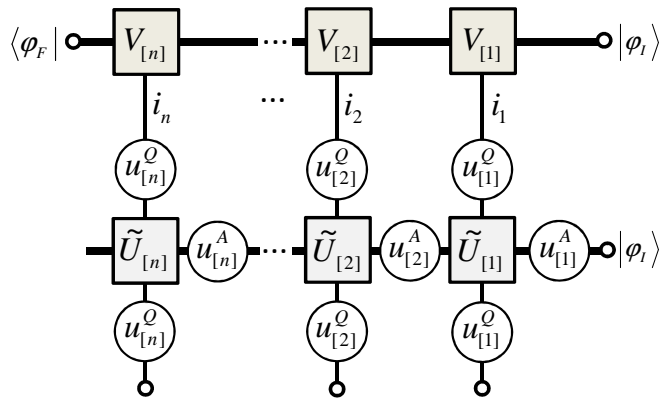}
\caption[MPS graphical representation of the overlap pattern for optimization of ancilla-qubit interactions]{MPS graphical representation of the overlap pattern for optimization of ancilla-qubit interactions upon incorporating the auxiliary single-qubit unitaries.}
\label{anc_qubit_ov_pat}
\end{figure}

However, we have found out that the fidelity can be improved upon systematically by allowing optimized rotations on the initial and final states of the qubits and ancilla, namely augmenting the restricted unitaries by unrestricted local \quot{auxiliary} operations $u^{\Q}$ and $u^{\A}$ as depicted schematically in Fig.~\ref{anc_qubit_ov_pat}. In this way, our numerics imply that a faithful representation of Gisin-Massar is obtained with a restricted experimentally realizable entangling Hamiltonian of $XXZ$ form up to $n=7$ qubits, as the second row of Table~\ref{table:AQ_reg} illustrates.


\section{Concluding remarks}
\label{sec:conclusions}

In conclusion, we have found strong numerical evidence that a state of Gisin-Massar form can be faithfully represented with an MPS of effective bond dimension $D=$3 up to $n=15$ qubits. Equivalently, it can be generated within sequential prescription with an ancillary system of experimentally manageable dimension $D=3$. We have also analyzed the possibility of realizing SQCM with a restricted class of physical interactions and have demonstrated that such restrictions do not pose any obstacle to sequential realization of cloning machine.

The current investigation and our results promises the possibility of elevating the SQCM ideas from an abstract theoretical level to practical recipes of experimental eminence.  They help to bridge the gap between theory and experiment
for realization of a sequential \quot{factory} of quantum cloning under real-life restrictions and in a form amenable to the existing physical setups. The achievements will be of importance, in particular, for realization of scalable quantum computing within all potentially scalable and hitherto realized QCMs, either in optical systems~\cite{Huang2001,Lamas2002,Martini2002,Fasel2002,Zhao2005,Bartuskova2007,Nagali2009} or NMR setups~\cite{Cummins2002,Du2005,Chen2007}. The method may accordingly prove useful for the recent experimental demonstration of \emph{probabilistic} quantum cloning~\cite{Chen2011} in an NMR quantum computer. Generalization to general sequential quantum cloning schemes for $N \to M$~\cite{Dang2008} may be also the subject of further investigation.





\begin{acknowledgments}
We gratefully acknowledge stimulating discussions with Enrique Solano, Lucas Lamata, and Andreas Weichselbaum. We particularly
would like to thank Andreas Weichselbaum for constructive comments on the manuscript.
H.S. acknowledges support from the vice-president office for research and technology affairs of Shahid Beheshti University, and
from Spintronics RTN, the DFG (SFB 631, De-730/3-2) and DFG under SFB 689, and the German
Excellence Initiative via the Nanosystems Initiative Munich (NIM).  H.S. is also grateful to
Universidad del Pa\'{\i}s Vasco for support and hospitality.

\end{acknowledgments}

\bibliography{SQC}

\begin{thebibliography}{48}%
\makeatletter
\providecommand \@ifxundefined [1]{%
 \@ifx{#1\undefined}
}%
\providecommand \@ifnum [1]{%
 \ifnum #1\expandafter \@firstoftwo
 \else \expandafter \@secondoftwo
 \fi
}%
\providecommand \@ifx [1]{%
 \ifx #1\expandafter \@firstoftwo
 \else \expandafter \@secondoftwo
 \fi
}%
\providecommand \natexlab [1]{#1}%
\providecommand \enquote  [1]{``#1''}%
\providecommand \bibnamefont  [1]{#1}%
\providecommand \bibfnamefont [1]{#1}%
\providecommand \citenamefont [1]{#1}%
\providecommand \href@noop [0]{\@secondoftwo}%
\providecommand \href [0]{\begingroup \@sanitize@url \@href}%
\providecommand \@href[1]{\@@startlink{#1}\@@href}%
\providecommand \@@href[1]{\endgroup#1\@@endlink}%
\providecommand \@sanitize@url [0]{\catcode `\\12\catcode `\$12\catcode
  `\&12\catcode `\#12\catcode `\^12\catcode `\_12\catcode `\%12\relax}%
\providecommand \@@startlink[1]{}%
\providecommand \@@endlink[0]{}%
\providecommand \url  [0]{\begingroup\@sanitize@url \@url }%
\providecommand \@url [1]{\endgroup\@href {#1}{\urlprefix }}%
\providecommand \urlprefix  [0]{URL }%
\providecommand \Eprint [0]{\href }%
\providecommand \doibase [0]{http://dx.doi.org/}%
\providecommand \selectlanguage [0]{\@gobble}%
\providecommand \bibinfo  [0]{\@secondoftwo}%
\providecommand \bibfield  [0]{\@secondoftwo}%
\providecommand \translation [1]{[#1]}%
\providecommand \BibitemOpen [0]{}%
\providecommand \bibitemStop [0]{}%
\providecommand \bibitemNoStop [0]{.\EOS\space}%
\providecommand \EOS [0]{\spacefactor3000\relax}%
\providecommand \BibitemShut  [1]{\csname bibitem#1\endcsname}%
\let\auto@bib@innerbib\@empty
\bibitem [{\citenamefont {Wootters}\ and\ \citenamefont
  {Zurek}(1982)}]{Wootters1982}%
  \BibitemOpen
  \bibfield  {author} {\bibinfo {author} {\bibfnamefont {W.~K.}\ \bibnamefont
  {Wootters}}\ and\ \bibinfo {author} {\bibfnamefont {W.~H.}\ \bibnamefont
  {Zurek}},\ }\href@noop {} {\bibfield  {journal} {\bibinfo  {journal}
  {Nature}\ }\textbf {\bibinfo {volume} {299}},\ \bibinfo {pages} {802}
  (\bibinfo {year} {1982})}\BibitemShut {NoStop}%
\bibitem [{\citenamefont {Nielsen}\ and\ \citenamefont
  {Chuang}(2000)}]{Nielsen2000}%
  \BibitemOpen
  \bibfield  {author} {\bibinfo {author} {\bibfnamefont {M.~A.}\ \bibnamefont
  {Nielsen}}\ and\ \bibinfo {author} {\bibfnamefont {I.~L.}\ \bibnamefont
  {Chuang}},\ }\href@noop {} {\emph {\bibinfo {title} {Quantum Computation and
  Quantum Information}}}\ (\bibinfo  {publisher} {Cambridge University Press,
  Cambridge},\ \bibinfo {year} {2000})\BibitemShut {NoStop}%
\bibitem [{\citenamefont {Bu\ifmmode~\check{z}\else \v{z}\fi{}ek}\ and\
  \citenamefont {Hillery}(1996)}]{Buzek1996}%
  \BibitemOpen
  \bibfield  {author} {\bibinfo {author} {\bibfnamefont {V.}~\bibnamefont
  {Bu\ifmmode~\check{z}\else \v{z}\fi{}ek}}\ and\ \bibinfo {author}
  {\bibfnamefont {M.}~\bibnamefont {Hillery}},\ }\href {\doibase
  10.1103/PhysRevA.54.1844} {\bibfield  {journal} {\bibinfo  {journal} {Phys.
  Rev. A}\ }\textbf {\bibinfo {volume} {54}},\ \bibinfo {pages} {1844}
  (\bibinfo {year} {1996})}\BibitemShut {NoStop}%
\bibitem [{\citenamefont {Gisin}\ and\ \citenamefont
  {Massar}(1997)}]{Gisin1997}%
  \BibitemOpen
  \bibfield  {author} {\bibinfo {author} {\bibfnamefont {N.}~\bibnamefont
  {Gisin}}\ and\ \bibinfo {author} {\bibfnamefont {S.}~\bibnamefont {Massar}},\
  }\href {\doibase 10.1103/PhysRevLett.79.2153} {\bibfield  {journal} {\bibinfo
   {journal} {Phys. Rev. Lett.}\ }\textbf {\bibinfo {volume} {79}},\ \bibinfo
  {pages} {2153} (\bibinfo {year} {1997})}\BibitemShut {NoStop}%
\bibitem [{\citenamefont {Scarani}\ \emph {et~al.}(2005)\citenamefont
  {Scarani}, \citenamefont {Iblisdir}, \citenamefont {Gisin},\ and\
  \citenamefont {Ac\'\i{}n}}]{Scarani2005}%
  \BibitemOpen
  \bibfield  {author} {\bibinfo {author} {\bibfnamefont {V.}~\bibnamefont
  {Scarani}}, \bibinfo {author} {\bibfnamefont {S.}~\bibnamefont {Iblisdir}},
  \bibinfo {author} {\bibfnamefont {N.}~\bibnamefont {Gisin}}, \ and\ \bibinfo
  {author} {\bibfnamefont {A.}~\bibnamefont {Ac\'\i{}n}},\ }\href {\doibase
  10.1103/RevModPhys.77.1225} {\bibfield  {journal} {\bibinfo  {journal} {Rev.
  Mod. Phys.}\ }\textbf {\bibinfo {volume} {77}},\ \bibinfo {pages} {1225}
  (\bibinfo {year} {2005})}\BibitemShut {NoStop}%
\bibitem [{\citenamefont {Huang}\ \emph {et~al.}(2001)\citenamefont {Huang},
  \citenamefont {Li}, \citenamefont {Li}, \citenamefont {Zhang}, \citenamefont
  {Jiang},\ and\ \citenamefont {Guo}}]{Huang2001}%
  \BibitemOpen
  \bibfield  {author} {\bibinfo {author} {\bibfnamefont {Y.-F.}\ \bibnamefont
  {Huang}}, \bibinfo {author} {\bibfnamefont {W.-L.}\ \bibnamefont {Li}},
  \bibinfo {author} {\bibfnamefont {C.-F.}\ \bibnamefont {Li}}, \bibinfo
  {author} {\bibfnamefont {Y.-S.}\ \bibnamefont {Zhang}}, \bibinfo {author}
  {\bibfnamefont {Y.-K.}\ \bibnamefont {Jiang}}, \ and\ \bibinfo {author}
  {\bibfnamefont {G.-C.}\ \bibnamefont {Guo}},\ }\href {\doibase
  10.1103/PhysRevA.64.012315} {\bibfield  {journal} {\bibinfo  {journal} {Phys.
  Rev. A}\ }\textbf {\bibinfo {volume} {64}},\ \bibinfo {pages} {012315}
  (\bibinfo {year} {2001})}\BibitemShut {NoStop}%
\bibitem [{\citenamefont {Lamas-Linares}\ \emph {et~al.}(2002)\citenamefont
  {Lamas-Linares}, \citenamefont {Simon}, \citenamefont {Howell},\ and\
  \citenamefont {Bouwmeester}}]{Lamas2002}%
  \BibitemOpen
  \bibfield  {author} {\bibinfo {author} {\bibfnamefont {A.}~\bibnamefont
  {Lamas-Linares}}, \bibinfo {author} {\bibfnamefont {C.}~\bibnamefont
  {Simon}}, \bibinfo {author} {\bibfnamefont {J.}~\bibnamefont {Howell}}, \
  and\ \bibinfo {author} {\bibfnamefont {D.}~\bibnamefont {Bouwmeester}},\
  }\href@noop {} {\bibfield  {journal} {\bibinfo  {journal} {Science}\ }\textbf
  {\bibinfo {volume} {296}},\ \bibinfo {pages} {712} (\bibinfo {year}
  {2002})}\BibitemShut {NoStop}%
\bibitem [{\citenamefont {De~Martini}\ \emph {et~al.}(2002)\citenamefont
  {De~Martini}, \citenamefont {Buzek}, \citenamefont {Sciarrino},\ and\
  \citenamefont {Sias}}]{Martini2002}%
  \BibitemOpen
  \bibfield  {author} {\bibinfo {author} {\bibfnamefont {F.}~\bibnamefont
  {De~Martini}}, \bibinfo {author} {\bibfnamefont {V.}~\bibnamefont {Buzek}},
  \bibinfo {author} {\bibfnamefont {F.}~\bibnamefont {Sciarrino}}, \ and\
  \bibinfo {author} {\bibfnamefont {C.}~\bibnamefont {Sias}},\ }\href@noop {}
  {\bibfield  {journal} {\bibinfo  {journal} {Nature}\ }\textbf {\bibinfo
  {volume} {419}},\ \bibinfo {pages} {815} (\bibinfo {year}
  {2002})}\BibitemShut {NoStop}%
\bibitem [{\citenamefont {Fasel}\ \emph {et~al.}(2002)\citenamefont {Fasel},
  \citenamefont {Gisin}, \citenamefont {Ribordy}, \citenamefont {Scarani},\
  and\ \citenamefont {Zbinden}}]{Fasel2002}%
  \BibitemOpen
  \bibfield  {author} {\bibinfo {author} {\bibfnamefont {S.}~\bibnamefont
  {Fasel}}, \bibinfo {author} {\bibfnamefont {N.}~\bibnamefont {Gisin}},
  \bibinfo {author} {\bibfnamefont {G.}~\bibnamefont {Ribordy}}, \bibinfo
  {author} {\bibfnamefont {V.}~\bibnamefont {Scarani}}, \ and\ \bibinfo
  {author} {\bibfnamefont {H.}~\bibnamefont {Zbinden}},\ }\href {\doibase
  10.1103/PhysRevLett.89.107901} {\bibfield  {journal} {\bibinfo  {journal}
  {Phys. Rev. Lett.}\ }\textbf {\bibinfo {volume} {89}},\ \bibinfo {pages}
  {107901} (\bibinfo {year} {2002})}\BibitemShut {NoStop}%
\bibitem [{\citenamefont {Zhao}\ \emph {et~al.}(2005)\citenamefont {Zhao},
  \citenamefont {Zhang}, \citenamefont {Zhou}, \citenamefont {Chen},
  \citenamefont {Lu}, \citenamefont {Karlsson},\ and\ \citenamefont
  {Pan}}]{Zhao2005}%
  \BibitemOpen
  \bibfield  {author} {\bibinfo {author} {\bibfnamefont {Z.}~\bibnamefont
  {Zhao}}, \bibinfo {author} {\bibfnamefont {A.-N.}\ \bibnamefont {Zhang}},
  \bibinfo {author} {\bibfnamefont {X.-Q.}\ \bibnamefont {Zhou}}, \bibinfo
  {author} {\bibfnamefont {Y.-A.}\ \bibnamefont {Chen}}, \bibinfo {author}
  {\bibfnamefont {C.-Y.}\ \bibnamefont {Lu}}, \bibinfo {author} {\bibfnamefont
  {A.}~\bibnamefont {Karlsson}}, \ and\ \bibinfo {author} {\bibfnamefont
  {J.-W.}\ \bibnamefont {Pan}},\ }\href {\doibase
  10.1103/PhysRevLett.95.030502} {\bibfield  {journal} {\bibinfo  {journal}
  {Phys. Rev. Lett.}\ }\textbf {\bibinfo {volume} {95}},\ \bibinfo {pages}
  {030502} (\bibinfo {year} {2005})}\BibitemShut {NoStop}%
\bibitem [{\citenamefont {Bart\ifmmode \mathring{u}\else
  \r{u}\fi{}\ifmmode~\check{s}\else \v{s}\fi{}kov\'a}\ \emph
  {et~al.}(2007)\citenamefont {Bart\ifmmode \mathring{u}\else
  \r{u}\fi{}\ifmmode~\check{s}\else \v{s}\fi{}kov\'a}, \citenamefont
  {Du\ifmmode~\check{s}\else \v{s}\fi{}ek}, \citenamefont
  {\ifmmode~\check{C}\else \v{C}\fi{}ernoch}, \citenamefont {Soubusta},\ and\
  \citenamefont {Fiur\'a\ifmmode~\check{s}\else
  \v{s}\fi{}ek}}]{Bartuskova2007}%
  \BibitemOpen
  \bibfield  {author} {\bibinfo {author} {\bibfnamefont {L.}~\bibnamefont
  {Bart\ifmmode \mathring{u}\else \r{u}\fi{}\ifmmode~\check{s}\else
  \v{s}\fi{}kov\'a}}, \bibinfo {author} {\bibfnamefont {M.}~\bibnamefont
  {Du\ifmmode~\check{s}\else \v{s}\fi{}ek}}, \bibinfo {author} {\bibfnamefont
  {A.}~\bibnamefont {\ifmmode~\check{C}\else \v{C}\fi{}ernoch}}, \bibinfo
  {author} {\bibfnamefont {J.}~\bibnamefont {Soubusta}}, \ and\ \bibinfo
  {author} {\bibfnamefont {J.}~\bibnamefont {Fiur\'a\ifmmode~\check{s}\else
  \v{s}\fi{}ek}},\ }\href {\doibase 10.1103/PhysRevLett.99.120505} {\bibfield
  {journal} {\bibinfo  {journal} {Phys. Rev. Lett.}\ }\textbf {\bibinfo
  {volume} {99}},\ \bibinfo {pages} {120505} (\bibinfo {year}
  {2007})}\BibitemShut {NoStop}%
\bibitem [{\citenamefont {Nagali}\ \emph {et~al.}(2009)\citenamefont {Nagali},
  \citenamefont {Sansoni}, \citenamefont {Sciarrino}, \citenamefont
  {De~Martini}, \citenamefont {Marrucci}, \citenamefont {Piccirillo},
  \citenamefont {Karimi},\ and\ \citenamefont {Santamato}}]{Nagali2009}%
  \BibitemOpen
  \bibfield  {author} {\bibinfo {author} {\bibfnamefont {E.}~\bibnamefont
  {Nagali}}, \bibinfo {author} {\bibfnamefont {L.}~\bibnamefont {Sansoni}},
  \bibinfo {author} {\bibfnamefont {F.}~\bibnamefont {Sciarrino}}, \bibinfo
  {author} {\bibfnamefont {F.}~\bibnamefont {De~Martini}}, \bibinfo {author}
  {\bibfnamefont {L.}~\bibnamefont {Marrucci}}, \bibinfo {author}
  {\bibfnamefont {B.}~\bibnamefont {Piccirillo}}, \bibinfo {author}
  {\bibfnamefont {E.}~\bibnamefont {Karimi}}, \ and\ \bibinfo {author}
  {\bibfnamefont {E.}~\bibnamefont {Santamato}},\ }\href@noop {} {\bibfield
  {journal} {\bibinfo  {journal} {Nat. Photon.}\ }\textbf {\bibinfo {volume}
  {3}},\ \bibinfo {pages} {720} (\bibinfo {year} {2009})}\BibitemShut {NoStop}%
\bibitem [{\citenamefont {Cummins}\ \emph {et~al.}(2002)\citenamefont
  {Cummins}, \citenamefont {Jones}, \citenamefont {Furze}, \citenamefont
  {Soffe}, \citenamefont {Mosca}, \citenamefont {Peach},\ and\ \citenamefont
  {Jones}}]{Cummins2002}%
  \BibitemOpen
  \bibfield  {author} {\bibinfo {author} {\bibfnamefont {H.~K.}\ \bibnamefont
  {Cummins}}, \bibinfo {author} {\bibfnamefont {C.}~\bibnamefont {Jones}},
  \bibinfo {author} {\bibfnamefont {A.}~\bibnamefont {Furze}}, \bibinfo
  {author} {\bibfnamefont {N.~F.}\ \bibnamefont {Soffe}}, \bibinfo {author}
  {\bibfnamefont {M.}~\bibnamefont {Mosca}}, \bibinfo {author} {\bibfnamefont
  {J.~M.}\ \bibnamefont {Peach}}, \ and\ \bibinfo {author} {\bibfnamefont
  {J.~A.}\ \bibnamefont {Jones}},\ }\href {\doibase
  10.1103/PhysRevLett.88.187901} {\bibfield  {journal} {\bibinfo  {journal}
  {Phys. Rev. Lett.}\ }\textbf {\bibinfo {volume} {88}},\ \bibinfo {pages}
  {187901} (\bibinfo {year} {2002})}\BibitemShut {NoStop}%
\bibitem [{\citenamefont {Du}\ \emph {et~al.}(2005)\citenamefont {Du},
  \citenamefont {Durt}, \citenamefont {Zou}, \citenamefont {Li}, \citenamefont
  {Kwek}, \citenamefont {Lai}, \citenamefont {Oh},\ and\ \citenamefont
  {Ekert}}]{Du2005}%
  \BibitemOpen
  \bibfield  {author} {\bibinfo {author} {\bibfnamefont {J.}~\bibnamefont
  {Du}}, \bibinfo {author} {\bibfnamefont {T.}~\bibnamefont {Durt}}, \bibinfo
  {author} {\bibfnamefont {P.}~\bibnamefont {Zou}}, \bibinfo {author}
  {\bibfnamefont {H.}~\bibnamefont {Li}}, \bibinfo {author} {\bibfnamefont
  {L.~C.}\ \bibnamefont {Kwek}}, \bibinfo {author} {\bibfnamefont {C.~H.}\
  \bibnamefont {Lai}}, \bibinfo {author} {\bibfnamefont {C.~H.}\ \bibnamefont
  {Oh}}, \ and\ \bibinfo {author} {\bibfnamefont {A.}~\bibnamefont {Ekert}},\
  }\href {\doibase 10.1103/PhysRevLett.94.040505} {\bibfield  {journal}
  {\bibinfo  {journal} {Phys. Rev. Lett.}\ }\textbf {\bibinfo {volume} {94}},\
  \bibinfo {pages} {040505} (\bibinfo {year} {2005})}\BibitemShut {NoStop}%
\bibitem [{\citenamefont {Chen}\ \emph {et~al.}(2007)\citenamefont {Chen},
  \citenamefont {Zhou}, \citenamefont {Suter},\ and\ \citenamefont
  {Du}}]{Chen2007}%
  \BibitemOpen
  \bibfield  {author} {\bibinfo {author} {\bibfnamefont {H.}~\bibnamefont
  {Chen}}, \bibinfo {author} {\bibfnamefont {X.}~\bibnamefont {Zhou}}, \bibinfo
  {author} {\bibfnamefont {D.}~\bibnamefont {Suter}}, \ and\ \bibinfo {author}
  {\bibfnamefont {J.}~\bibnamefont {Du}},\ }\href {\doibase
  10.1103/PhysRevA.75.012317} {\bibfield  {journal} {\bibinfo  {journal} {Phys.
  Rev. A}\ }\textbf {\bibinfo {volume} {75}},\ \bibinfo {pages} {012317}
  (\bibinfo {year} {2007})}\BibitemShut {NoStop}%
\bibitem [{\citenamefont {Delgado}\ \emph {et~al.}(2007)\citenamefont
  {Delgado}, \citenamefont {Lamata}, \citenamefont {Le\'on}, \citenamefont
  {Salgado},\ and\ \citenamefont {Solano}}]{Delgado2007}%
  \BibitemOpen
  \bibfield  {author} {\bibinfo {author} {\bibfnamefont {Y.}~\bibnamefont
  {Delgado}}, \bibinfo {author} {\bibfnamefont {L.}~\bibnamefont {Lamata}},
  \bibinfo {author} {\bibfnamefont {J.}~\bibnamefont {Le\'on}}, \bibinfo
  {author} {\bibfnamefont {D.}~\bibnamefont {Salgado}}, \ and\ \bibinfo
  {author} {\bibfnamefont {E.}~\bibnamefont {Solano}},\ }\href {\doibase
  10.1103/PhysRevLett.98.150502} {\bibfield  {journal} {\bibinfo  {journal}
  {Phys. Rev. Lett.}\ }\textbf {\bibinfo {volume} {98}},\ \bibinfo {pages}
  {150502} (\bibinfo {year} {2007})}\BibitemShut {NoStop}%
\bibitem [{\citenamefont {Lamata}\ \emph {et~al.}(2008)\citenamefont {Lamata},
  \citenamefont {Le\'on}, \citenamefont {P\'erez-Garc\'\i{}a}, \citenamefont
  {Salgado},\ and\ \citenamefont {Solano}}]{Lamata2008}%
  \BibitemOpen
  \bibfield  {author} {\bibinfo {author} {\bibfnamefont {L.}~\bibnamefont
  {Lamata}}, \bibinfo {author} {\bibfnamefont {J.}~\bibnamefont {Le\'on}},
  \bibinfo {author} {\bibfnamefont {D.}~\bibnamefont {P\'erez-Garc\'\i{}a}},
  \bibinfo {author} {\bibfnamefont {D.}~\bibnamefont {Salgado}}, \ and\
  \bibinfo {author} {\bibfnamefont {E.}~\bibnamefont {Solano}},\ }\href
  {\doibase 10.1103/PhysRevLett.101.180506} {\bibfield  {journal} {\bibinfo
  {journal} {Phys. Rev. Lett.}\ }\textbf {\bibinfo {volume} {101}},\ \bibinfo
  {pages} {180506} (\bibinfo {year} {2008})}\BibitemShut {NoStop}%
\bibitem [{\citenamefont {Saberi}(2011)}]{Saberi2011}%
  \BibitemOpen
  \bibfield  {author} {\bibinfo {author} {\bibfnamefont {H.}~\bibnamefont
  {Saberi}},\ }\href {\doibase 10.1103/PhysRevA.84.032323} {\bibfield
  {journal} {\bibinfo  {journal} {Phys. Rev. A}\ }\textbf {\bibinfo {volume}
  {84}},\ \bibinfo {pages} {032323} (\bibinfo {year} {2011})}\BibitemShut
  {NoStop}%
\bibitem [{\citenamefont {Kuhn}\ \emph {et~al.}(2002)\citenamefont {Kuhn},
  \citenamefont {Hennrich},\ and\ \citenamefont {Rempe}}]{Kuhn2002}%
  \BibitemOpen
  \bibfield  {author} {\bibinfo {author} {\bibfnamefont {A.}~\bibnamefont
  {Kuhn}}, \bibinfo {author} {\bibfnamefont {M.}~\bibnamefont {Hennrich}}, \
  and\ \bibinfo {author} {\bibfnamefont {G.}~\bibnamefont {Rempe}},\ }\href
  {\doibase 10.1103/PhysRevLett.89.067901} {\bibfield  {journal} {\bibinfo
  {journal} {Phys. Rev. Lett.}\ }\textbf {\bibinfo {volume} {89}},\ \bibinfo
  {pages} {067901} (\bibinfo {year} {2002})}\BibitemShut {NoStop}%
\bibitem [{\citenamefont {McKeever}\ \emph {et~al.}(2004)\citenamefont
  {McKeever}, \citenamefont {Boca}, \citenamefont {Boozer}, \citenamefont
  {Miller}, \citenamefont {Buck}, \citenamefont {Kuzmich},\ and\ \citenamefont
  {Kimble}}]{McKeever2004}%
  \BibitemOpen
  \bibfield  {author} {\bibinfo {author} {\bibfnamefont {J.}~\bibnamefont
  {McKeever}}, \bibinfo {author} {\bibfnamefont {A.}~\bibnamefont {Boca}},
  \bibinfo {author} {\bibfnamefont {A.}~\bibnamefont {Boozer}}, \bibinfo
  {author} {\bibfnamefont {R.}~\bibnamefont {Miller}}, \bibinfo {author}
  {\bibfnamefont {J.}~\bibnamefont {Buck}}, \bibinfo {author} {\bibfnamefont
  {A.}~\bibnamefont {Kuzmich}}, \ and\ \bibinfo {author} {\bibfnamefont
  {H.}~\bibnamefont {Kimble}},\ }\href@noop {} {\bibfield  {journal} {\bibinfo
  {journal} {Science}\ }\textbf {\bibinfo {volume} {303}},\ \bibinfo {pages}
  {1992} (\bibinfo {year} {2004})}\BibitemShut {NoStop}%
\bibitem [{\citenamefont {Sch\"on}\ \emph {et~al.}(2005)\citenamefont
  {Sch\"on}, \citenamefont {Solano}, \citenamefont {Verstraete}, \citenamefont
  {Cirac},\ and\ \citenamefont {Wolf}}]{Schoen2005}%
  \BibitemOpen
  \bibfield  {author} {\bibinfo {author} {\bibfnamefont {C.}~\bibnamefont
  {Sch\"on}}, \bibinfo {author} {\bibfnamefont {E.}~\bibnamefont {Solano}},
  \bibinfo {author} {\bibfnamefont {F.}~\bibnamefont {Verstraete}}, \bibinfo
  {author} {\bibfnamefont {J.~I.}\ \bibnamefont {Cirac}}, \ and\ \bibinfo
  {author} {\bibfnamefont {M.~M.}\ \bibnamefont {Wolf}},\ }\href {\doibase
  10.1103/PhysRevLett.95.110503} {\bibfield  {journal} {\bibinfo  {journal}
  {Phys. Rev. Lett.}\ }\textbf {\bibinfo {volume} {95}},\ \bibinfo {eid}
  {110503} (\bibinfo {year} {2005})}\BibitemShut {NoStop}%
\bibitem [{\citenamefont {Sch\"on}(2005)}]{Schoen2005_2}%
  \BibitemOpen
  \bibfield  {author} {\bibinfo {author} {\bibfnamefont {C.}~\bibnamefont
  {Sch\"on}},\ }\href@noop {} {Ph.D. thesis},\ \bibinfo  {school} {Technical
  University of Munich} (\bibinfo {year} {2005})\BibitemShut {NoStop}%
\bibitem [{\citenamefont {Sch\"on}\ \emph {et~al.}(2007)\citenamefont
  {Sch\"on}, \citenamefont {Hammerer}, \citenamefont {Wolf}, \citenamefont
  {Cirac},\ and\ \citenamefont {Solano}}]{Schoen2007}%
  \BibitemOpen
  \bibfield  {author} {\bibinfo {author} {\bibfnamefont {C.}~\bibnamefont
  {Sch\"on}}, \bibinfo {author} {\bibfnamefont {K.}~\bibnamefont {Hammerer}},
  \bibinfo {author} {\bibfnamefont {M.~M.}\ \bibnamefont {Wolf}}, \bibinfo
  {author} {\bibfnamefont {J.~I.}\ \bibnamefont {Cirac}}, \ and\ \bibinfo
  {author} {\bibfnamefont {E.}~\bibnamefont {Solano}},\ }\href {\doibase
  10.1103/PhysRevA.75.032311} {\bibfield  {journal} {\bibinfo  {journal} {Phys.
  Rev. A}\ }\textbf {\bibinfo {volume} {75}},\ \bibinfo {eid} {032311}
  (\bibinfo {year} {2007})}\BibitemShut {NoStop}%
\bibitem [{\citenamefont {Wilk}\ \emph {et~al.}(2007)\citenamefont {Wilk},
  \citenamefont {Webster}, \citenamefont {Kuhn},\ and\ \citenamefont
  {Rempe}}]{Wilk2007}%
  \BibitemOpen
  \bibfield  {author} {\bibinfo {author} {\bibfnamefont {T.}~\bibnamefont
  {Wilk}}, \bibinfo {author} {\bibfnamefont {S.}~\bibnamefont {Webster}},
  \bibinfo {author} {\bibfnamefont {A.}~\bibnamefont {Kuhn}}, \ and\ \bibinfo
  {author} {\bibfnamefont {G.}~\bibnamefont {Rempe}},\ }\href@noop {}
  {\bibfield  {journal} {\bibinfo  {journal} {Science}\ }\textbf {\bibinfo
  {volume} {317}},\ \bibinfo {pages} {488} (\bibinfo {year}
  {2007})}\BibitemShut {NoStop}%
\bibitem [{\citenamefont {Weber}\ \emph {et~al.}(2009)\citenamefont {Weber},
  \citenamefont {Specht}, \citenamefont {M\"uller}, \citenamefont {Bochmann},
  \citenamefont {M\"ucke}, \citenamefont {Moehring},\ and\ \citenamefont
  {Rempe}}]{Weber2009}%
  \BibitemOpen
  \bibfield  {author} {\bibinfo {author} {\bibfnamefont {B.}~\bibnamefont
  {Weber}}, \bibinfo {author} {\bibfnamefont {H.~P.}\ \bibnamefont {Specht}},
  \bibinfo {author} {\bibfnamefont {T.}~\bibnamefont {M\"uller}}, \bibinfo
  {author} {\bibfnamefont {J.}~\bibnamefont {Bochmann}}, \bibinfo {author}
  {\bibfnamefont {M.}~\bibnamefont {M\"ucke}}, \bibinfo {author} {\bibfnamefont
  {D.~L.}\ \bibnamefont {Moehring}}, \ and\ \bibinfo {author} {\bibfnamefont
  {G.}~\bibnamefont {Rempe}},\ }\href {\doibase 10.1103/PhysRevLett.102.030501}
  {\bibfield  {journal} {\bibinfo  {journal} {Phys. Rev. Lett.}\ }\textbf
  {\bibinfo {volume} {102}},\ \bibinfo {pages} {030501} (\bibinfo {year}
  {2009})}\BibitemShut {NoStop}%
\bibitem [{\citenamefont {Saberi}\ \emph {et~al.}(2009)\citenamefont {Saberi},
  \citenamefont {Weichselbaum}, \citenamefont {Lamata}, \citenamefont
  {P\'erez-Garc\'\i{}a}, \citenamefont {von Delft},\ and\ \citenamefont
  {Solano}}]{Saberi2009}%
  \BibitemOpen
  \bibfield  {author} {\bibinfo {author} {\bibfnamefont {H.}~\bibnamefont
  {Saberi}}, \bibinfo {author} {\bibfnamefont {A.}~\bibnamefont
  {Weichselbaum}}, \bibinfo {author} {\bibfnamefont {L.}~\bibnamefont
  {Lamata}}, \bibinfo {author} {\bibfnamefont {D.}~\bibnamefont
  {P\'erez-Garc\'\i{}a}}, \bibinfo {author} {\bibfnamefont {J.}~\bibnamefont
  {von Delft}}, \ and\ \bibinfo {author} {\bibfnamefont {E.}~\bibnamefont
  {Solano}},\ }\href {\doibase 10.1103/PhysRevA.80.022334} {\bibfield
  {journal} {\bibinfo  {journal} {Phys. Rev. A}\ }\textbf {\bibinfo {volume}
  {80}},\ \bibinfo {pages} {022334} (\bibinfo {year} {2009})}\BibitemShut
  {NoStop}%
\bibitem [{\citenamefont {Christ}\ \emph {et~al.}(2008)\citenamefont {Christ},
  \citenamefont {Cirac},\ and\ \citenamefont {Giedke}}]{Christ2008}%
  \BibitemOpen
  \bibfield  {author} {\bibinfo {author} {\bibfnamefont {H.}~\bibnamefont
  {Christ}}, \bibinfo {author} {\bibfnamefont {J.~I.}\ \bibnamefont {Cirac}}, \
  and\ \bibinfo {author} {\bibfnamefont {G.}~\bibnamefont {Giedke}},\ }\href
  {\doibase 10.1103/PhysRevB.78.125314} {\bibfield  {journal} {\bibinfo
  {journal} {Phys. Rev. B}\ }\textbf {\bibinfo {volume} {78}},\ \bibinfo
  {pages} {125314} (\bibinfo {year} {2008})}\BibitemShut {NoStop}%
\bibitem [{\citenamefont {Perez-Garcia}\ \emph {et~al.}(2007)\citenamefont
  {Perez-Garcia}, \citenamefont {Verstraete}, \citenamefont {Wolf},\ and\
  \citenamefont {Cirac}}]{Perez2007}%
  \BibitemOpen
  \bibfield  {author} {\bibinfo {author} {\bibfnamefont {D.}~\bibnamefont
  {Perez-Garcia}}, \bibinfo {author} {\bibfnamefont {F.}~\bibnamefont
  {Verstraete}}, \bibinfo {author} {\bibfnamefont {M.~M.}\ \bibnamefont
  {Wolf}}, \ and\ \bibinfo {author} {\bibfnamefont {J.~I.}\ \bibnamefont
  {Cirac}},\ }\href
  {http://www.citebase.org/abstract?id=oai:arXiv.org:quant-ph/0608197}
  {\bibfield  {journal} {\bibinfo  {journal} {Quantum Inf. Comput.}\ }\textbf
  {\bibinfo {volume} {7}},\ \bibinfo {pages} {401} (\bibinfo {year}
  {2007})}\BibitemShut {NoStop}%
\bibitem [{\citenamefont {Fannes}\ \emph {et~al.}(1992)\citenamefont {Fannes},
  \citenamefont {Nachtergaele},\ and\ \citenamefont {Werner}}]{Fannes1992}%
  \BibitemOpen
  \bibfield  {author} {\bibinfo {author} {\bibfnamefont {M.}~\bibnamefont
  {Fannes}}, \bibinfo {author} {\bibfnamefont {B.}~\bibnamefont
  {Nachtergaele}}, \ and\ \bibinfo {author} {\bibfnamefont {R.}~\bibnamefont
  {Werner}},\ }\href {\doibase 10.1007/BF02099178} {\bibfield  {journal}
  {\bibinfo  {journal} {Commun. Math. Phys.}\ }\textbf {\bibinfo {volume}
  {144}},\ \bibinfo {pages} {443} (\bibinfo {year} {1992})}\BibitemShut
  {NoStop}%
\bibitem [{\citenamefont {Saberi}\ \emph {et~al.}(2008)\citenamefont {Saberi},
  \citenamefont {Weichselbaum},\ and\ \citenamefont {von Delft}}]{Saberi2008}%
  \BibitemOpen
  \bibfield  {author} {\bibinfo {author} {\bibfnamefont {H.}~\bibnamefont
  {Saberi}}, \bibinfo {author} {\bibfnamefont {A.}~\bibnamefont
  {Weichselbaum}}, \ and\ \bibinfo {author} {\bibfnamefont {J.}~\bibnamefont
  {von Delft}},\ }\href {\doibase 10.1103/PhysRevB.78.035124} {\bibfield
  {journal} {\bibinfo  {journal} {Phys. Rev. B}\ }\textbf {\bibinfo {volume}
  {78}},\ \bibinfo {eid} {035124} (\bibinfo {year} {2008})}\BibitemShut
  {NoStop}%
\bibitem [{\citenamefont {Weichselbaum}\ \emph {et~al.}(2009)\citenamefont
  {Weichselbaum}, \citenamefont {Verstraete}, \citenamefont {Schollw\"ock},
  \citenamefont {Cirac},\ and\ \citenamefont {von Delft}}]{Weichselbaum2009}%
  \BibitemOpen
  \bibfield  {author} {\bibinfo {author} {\bibfnamefont {A.}~\bibnamefont
  {Weichselbaum}}, \bibinfo {author} {\bibfnamefont {F.}~\bibnamefont
  {Verstraete}}, \bibinfo {author} {\bibfnamefont {U.}~\bibnamefont
  {Schollw\"ock}}, \bibinfo {author} {\bibfnamefont {J.~I.}\ \bibnamefont
  {Cirac}}, \ and\ \bibinfo {author} {\bibfnamefont {J.}~\bibnamefont {von
  Delft}},\ }\href {\doibase 10.1103/PhysRevB.80.165117} {\bibfield  {journal}
  {\bibinfo  {journal} {Phys. Rev. B}\ }\textbf {\bibinfo {volume} {80}},\
  \bibinfo {pages} {165117} (\bibinfo {year} {2009})}\BibitemShut {NoStop}%
\bibitem [{\citenamefont {Pi\ifmmode~\check{z}\else \v{z}\fi{}orn}\ and\
  \citenamefont {Verstraete}(2012)}]{Pizorn2012}%
  \BibitemOpen
  \bibfield  {author} {\bibinfo {author} {\bibfnamefont {I.}~\bibnamefont
  {Pi\ifmmode~\check{z}\else \v{z}\fi{}orn}}\ and\ \bibinfo {author}
  {\bibfnamefont {F.}~\bibnamefont {Verstraete}},\ }\href {\doibase
  10.1103/PhysRevLett.108.067202} {\bibfield  {journal} {\bibinfo  {journal}
  {Phys. Rev. Lett.}\ }\textbf {\bibinfo {volume} {108}},\ \bibinfo {pages}
  {067202} (\bibinfo {year} {2012})}\BibitemShut {NoStop}%
\bibitem [{\citenamefont {Vidal}(2003)}]{Vidal2003}%
  \BibitemOpen
  \bibfield  {author} {\bibinfo {author} {\bibfnamefont {G.}~\bibnamefont
  {Vidal}},\ }\href {\doibase 10.1103/PhysRevLett.91.147902} {\bibfield
  {journal} {\bibinfo  {journal} {Phys. Rev. Lett.}\ }\textbf {\bibinfo
  {volume} {91}},\ \bibinfo {pages} {147902} (\bibinfo {year}
  {2003})}\BibitemShut {NoStop}%
\bibitem [{\citenamefont {Golub}\ and\ \citenamefont {Loan}(1996)}]{Golub1996}%
  \BibitemOpen
  \bibfield  {author} {\bibinfo {author} {\bibfnamefont {G.~H.}\ \bibnamefont
  {Golub}}\ and\ \bibinfo {author} {\bibfnamefont {C.~F.~V.}\ \bibnamefont
  {Loan}},\ }\href@noop {} {\emph {\bibinfo {title} {Matrix Computations}}}\
  (\bibinfo  {publisher} {The John Hopkins University Press, Baltimore},\
  \bibinfo {year} {1996})\BibitemShut {NoStop}%
\bibitem [{\citenamefont {Horn}\ and\ \citenamefont
  {Johnson}(1991)}]{Horn1991}%
  \BibitemOpen
  \bibfield  {author} {\bibinfo {author} {\bibfnamefont {R.~A.}\ \bibnamefont
  {Horn}}\ and\ \bibinfo {author} {\bibfnamefont {C.~R.}\ \bibnamefont
  {Johnson}},\ }\href@noop {} {\emph {\bibinfo {title} {Topics in Matrix
  Analysis}}}\ (\bibinfo  {publisher} {Cambridge University Press, Cambridge},\
  \bibinfo {year} {1991})\BibitemShut {NoStop}%
\bibitem [{\citenamefont {Bennett}\ \emph {et~al.}(1996)\citenamefont
  {Bennett}, \citenamefont {Mor},\ and\ \citenamefont {Smolin}}]{Bennett1996}%
  \BibitemOpen
  \bibfield  {author} {\bibinfo {author} {\bibfnamefont {C.~H.}\ \bibnamefont
  {Bennett}}, \bibinfo {author} {\bibfnamefont {T.}~\bibnamefont {Mor}}, \ and\
  \bibinfo {author} {\bibfnamefont {J.~A.}\ \bibnamefont {Smolin}},\ }\href
  {\doibase 10.1103/PhysRevA.54.2675} {\bibfield  {journal} {\bibinfo
  {journal} {Phys. Rev. A}\ }\textbf {\bibinfo {volume} {54}},\ \bibinfo
  {pages} {2675} (\bibinfo {year} {1996})}\BibitemShut {NoStop}%
\bibitem [{\citenamefont {Hayes}\ \emph {et~al.}(2010)\citenamefont {Hayes},
  \citenamefont {Haselgrove}, \citenamefont {Gilchrist},\ and\ \citenamefont
  {Ralph}}]{Hayes2010}%
  \BibitemOpen
  \bibfield  {author} {\bibinfo {author} {\bibfnamefont {A.~J.~F.}\
  \bibnamefont {Hayes}}, \bibinfo {author} {\bibfnamefont {H.~L.}\ \bibnamefont
  {Haselgrove}}, \bibinfo {author} {\bibfnamefont {A.}~\bibnamefont
  {Gilchrist}}, \ and\ \bibinfo {author} {\bibfnamefont {T.~C.}\ \bibnamefont
  {Ralph}},\ }\href {\doibase 10.1103/PhysRevA.82.022323} {\bibfield  {journal}
  {\bibinfo  {journal} {Phys. Rev. A}\ }\textbf {\bibinfo {volume} {82}},\
  \bibinfo {pages} {022323} (\bibinfo {year} {2010})}\BibitemShut {NoStop}%
\bibitem [{\citenamefont {Saberi}(2009)}]{Saberi2009_2}%
  \BibitemOpen
  \bibfield  {author} {\bibinfo {author} {\bibfnamefont {H.}~\bibnamefont
  {Saberi}},\ }\href@noop {} {Ph.D. thesis},\ \bibinfo  {school}
  {Ludwig-Maximilians University, Munich} (\bibinfo {year} {2009})\BibitemShut
  {NoStop}%
\bibitem [{\citenamefont {Andrews}\ and\ \citenamefont
  {Patterson}(1976)}]{Andrews1976}%
  \BibitemOpen
  \bibfield  {author} {\bibinfo {author} {\bibfnamefont {H.}~\bibnamefont
  {Andrews}}\ and\ \bibinfo {author} {\bibfnamefont {C.}~\bibnamefont
  {Patterson}},\ }\href@noop {} {\bibfield  {journal} {\bibinfo  {journal}
  {Communications, IEEE Transactions on [legacy, pre - 1988]}\ }\textbf
  {\bibinfo {volume} {24}},\ \bibinfo {pages} {425} (\bibinfo {year}
  {1976})}\BibitemShut {NoStop}%
\bibitem [{fro()}]{frob}%
  \BibitemOpen
  \href@noop {} {}\bibinfo {note} {The Frobenius or Hilbert-Schmidt norm of an
  operator is defined by ${\|O\|}_F=\sqrt{\mathrm{Tr}(O^{\dagger} \:
  O)}$.}\BibitemShut {Stop}%
\bibitem [{\citenamefont {Verstraete}\ \emph {et~al.}(2004)\citenamefont
  {Verstraete}, \citenamefont {Porras},\ and\ \citenamefont
  {Cirac}}]{Verstraete2004_2}%
  \BibitemOpen
  \bibfield  {author} {\bibinfo {author} {\bibfnamefont {F.}~\bibnamefont
  {Verstraete}}, \bibinfo {author} {\bibfnamefont {D.}~\bibnamefont {Porras}},
  \ and\ \bibinfo {author} {\bibfnamefont {J.~I.}\ \bibnamefont {Cirac}},\
  }\href {\doibase 10.1103/PhysRevLett.93.227205} {\bibfield  {journal}
  {\bibinfo  {journal} {Phys. Rev. Lett.}\ }\textbf {\bibinfo {volume} {93}},\
  \bibinfo {pages} {227205} (\bibinfo {year} {2004})}\BibitemShut {NoStop}%
\bibitem [{\citenamefont {Schollw\"{o}ck}(2005)}]{Schollwock2005}%
  \BibitemOpen
  \bibfield  {author} {\bibinfo {author} {\bibfnamefont {U.}~\bibnamefont
  {Schollw\"{o}ck}},\ }\href {\doibase 10.1103/RevModPhys.77.259} {\bibfield
  {journal} {\bibinfo  {journal} {Rev. Mod. Phys.}\ }\textbf {\bibinfo {volume}
  {77}},\ \bibinfo {eid} {259} (\bibinfo {year} {2005})}\BibitemShut {NoStop}%
\bibitem [{\citenamefont {White}(1992)}]{White1992}%
  \BibitemOpen
  \bibfield  {author} {\bibinfo {author} {\bibfnamefont {S.~R.}\ \bibnamefont
  {White}},\ }\href {\doibase 10.1103/PhysRevLett.69.2863} {\bibfield
  {journal} {\bibinfo  {journal} {Phys. Rev. Lett.}\ }\textbf {\bibinfo
  {volume} {69}},\ \bibinfo {pages} {2863} (\bibinfo {year}
  {1992})}\BibitemShut {NoStop}%
\bibitem [{\citenamefont {Bertsekas}(1996)}]{Bertsekas1996}%
  \BibitemOpen
  \bibfield  {author} {\bibinfo {author} {\bibfnamefont {D.~P.}\ \bibnamefont
  {Bertsekas}},\ }\href@noop {} {\emph {\bibinfo {title} {Constrained
  Optimization and Lagrange Multiplier Methods}}}\ (\bibinfo  {publisher}
  {Athena Scientific, Belmont},\ \bibinfo {year} {1996})\BibitemShut {NoStop}%
\bibitem [{\citenamefont {Brown}\ \emph {et~al.}(2003)\citenamefont {Brown},
  \citenamefont {Vala},\ and\ \citenamefont {Whaley}}]{Brown2003}%
  \BibitemOpen
  \bibfield  {author} {\bibinfo {author} {\bibfnamefont {K.~R.}\ \bibnamefont
  {Brown}}, \bibinfo {author} {\bibfnamefont {J.}~\bibnamefont {Vala}}, \ and\
  \bibinfo {author} {\bibfnamefont {K.~B.}\ \bibnamefont {Whaley}},\ }\href
  {\doibase 10.1103/PhysRevA.67.012309} {\bibfield  {journal} {\bibinfo
  {journal} {Phys. Rev. A}\ }\textbf {\bibinfo {volume} {67}},\ \bibinfo
  {pages} {012309} (\bibinfo {year} {2003})}\BibitemShut {NoStop}%
\bibitem [{uni()}]{uni_iso}%
  \BibitemOpen
  \href@noop {} {}\bibinfo {note} {Note that the action of each restricted
  unitary on an input qubit, $\tilde{U}_{[k]}^{\A\Q} |\psi\rangle$, produces a
  restricted isometry of the form $\sum_{i_k,j_k,\alpha,\beta}
  \tilde{U}_{\alpha,\beta}^{i_k,j_k} |\alpha i_k\rangle \langle \beta j_k|
  \psi\rangle=\sum_{i_k,\alpha,\beta} \tilde{V}_{\alpha,\beta}^{i_k}|\alpha
  i_k\rangle \langle \beta|$ with the definition
  $\tilde{V}_{\alpha,\beta}^{i_k} \equiv \sum_{j_k}
  \tilde{U}_{\alpha,\beta}^{i_k,j_k} \langle j_k| \psi\rangle$ for the
  resulting isometry $\tilde{V}_{[k]}^{\A\Q}$.}\BibitemShut {Stop}%
\bibitem [{\citenamefont {Chen}\ \emph {et~al.}(2011)\citenamefont {Chen},
  \citenamefont {Lu}, \citenamefont {Chong}, \citenamefont {Qin}, \citenamefont
  {Zhou}, \citenamefont {Peng},\ and\ \citenamefont {Du}}]{Chen2011}%
  \BibitemOpen
  \bibfield  {author} {\bibinfo {author} {\bibfnamefont {H.}~\bibnamefont
  {Chen}}, \bibinfo {author} {\bibfnamefont {D.}~\bibnamefont {Lu}}, \bibinfo
  {author} {\bibfnamefont {B.}~\bibnamefont {Chong}}, \bibinfo {author}
  {\bibfnamefont {G.}~\bibnamefont {Qin}}, \bibinfo {author} {\bibfnamefont
  {X.}~\bibnamefont {Zhou}}, \bibinfo {author} {\bibfnamefont {X.}~\bibnamefont
  {Peng}}, \ and\ \bibinfo {author} {\bibfnamefont {J.}~\bibnamefont {Du}},\
  }\href {\doibase 10.1103/PhysRevLett.106.180404} {\bibfield  {journal}
  {\bibinfo  {journal} {Phys. Rev. Lett.}\ }\textbf {\bibinfo {volume} {106}},\
  \bibinfo {pages} {180404} (\bibinfo {year} {2011})}\BibitemShut {NoStop}%
\bibitem [{\citenamefont {Dang}\ and\ \citenamefont {Fan}(2008)}]{Dang2008}%
  \BibitemOpen
  \bibfield  {author} {\bibinfo {author} {\bibfnamefont {G.-F.}\ \bibnamefont
  {Dang}}\ and\ \bibinfo {author} {\bibfnamefont {H.}~\bibnamefont {Fan}},\
  }\href@noop {} {\bibfield  {journal} {\bibinfo  {journal} {J. Phys. A: Math.
  Theor.}\ }\textbf {\bibinfo {volume} {41}},\ \bibinfo {pages} {155303}
  (\bibinfo {year} {2008})}\BibitemShut {NoStop}%
\end{thebibliography}%

\end{document}